\pdfoutput=1
\documentclass[fleqn,usenatbib,useAMS]{mnras}
\usepackage{newtxtext,newtxmath}
\usepackage{amsmath,amssymb}
\usepackage[T1]{fontenc}
\usepackage{natbib}
\usepackage{epsfig}
\usepackage{graphicx}
\usepackage{tablefootnote}
\usepackage{floatrow}
\usepackage{caption}

\title[Projected bounds on ALPs from ATHENA]{Projected bounds on ALPs from {\it Athena}}
\author[J. P. Conlon et al.]{
Joseph P. Conlon$^{1}$,
Francesca Day$^{1}$, 
Nicholas Jennings$^{1}$\thanks{E-mail: nicholas.jennings@physics.ox.ac.uk}, 
Sven Krippendorf$^{1}$
\newauthor{ and Francesco Muia$^{1}$}
\\
$^{1}$Rudolf Peierls Centre for Theoretical Physics, 1 Keble Road, Oxford, OX1 3NP, UK\\
}
\date{\today}

\pubyear{2017}

\begin{document}
\label{firstpage}
\pagerange{\pageref{firstpage}--\pageref{lastpage}}
\maketitle

\begin{abstract}
Galaxy clusters represent excellent laboratories to search for Axion-Like Particles (ALPs). They contain magnetic fields which can induce quasi-sinusoidal oscillations in the X-ray spectra of AGNs situated in or behind them. Due to its excellent 
energy resolution, the X-ray Integral Field Unit (X-IFU) instrument onboard the {\it Athena} X-ray Observatory will be far more sensitive to ALP-induced modulations than current detectors. As a first analysis of the sensitivity of {\it Athena} to the ALP-photon coupling $g_{a \gamma \gamma}$, we simulate observations of the Seyfert galaxy NGC~1275 (hosting the radio source 3C~84) in the Perseus cluster using the SIXTE simulation software. We estimate that for a 200~ks exposure, a non-observation of spectral modulations will constrain ${g_{a\gamma\gamma}\lesssim1.5\times10^{-13}~\rm{GeV}^{-1}}$ for $m_a\lesssim10^{-12}~\rm{eV}$, representing an order of magnitude improvement over constraints derived  using the current generation of satellites.
\end{abstract}


\section{Introduction}
\label{introduction}
X-ray astronomy provides a novel arena for fundamental physics. Thanks to exciting recent data, such as the observed excess at 3.5~keV \citep{Bulbul, Boyarsky}, there has been a renewed interest among particle physicists in the great promise of X-ray astronomy to shed light on physics beyond the Standard Model, including the existence of new particles.

One area for which X-ray astronomy is particularly suitable
 is in the search for Axion-Like Particles (ALPs).
ALPs are light pseudo-scalars that are a well motivated extension of the Standard Model \citep{PecceiQuinn,Wilczek,Weinberg} that arise
 generically in string compactifications, for example see \citep{hep-th/0602233,hep-th/0605206,1206.0819}. A general review of ALPs is \citep{RingwaldReview}. In the presence of a magnetic field $\langle B \rangle$ ALPs and photons interconvert \citep{Sikivie:1983ip, Raffelt:1987im}, and this induces quasi-sinusoidal oscillations at X-ray energies in the spectra of sources in and around galaxy clusters \citep{1304.0989, 1509.06748}.

Searches for these oscillations can be used to constrain ALP parameter space.
Current constraints on ALPs derived in this fashion \citep{1304.0989, Berg:2016ese, Marsh:2017yvc,Conlon:2017qcw} are based on data taken with CCD detectors, which have an energy resolution of $\mathcal{O}(100~\rm{eV})$. A large improvement with sensitivity will be achieved once data becomes available from microcalorimeters with $\mathcal{O}({\rm a~few} \, \rm{eV})$ energy resolution.
Such microcalorimeters will be on board the Advanced Telescope for High ENergy Astrophysics (ATHENA), currently scheduled to launch in 2028.
Its X-IFU instrument will have both large effective area, good imaging and energy resolution of $\sim 2.5 \rm{eV}$, greatly enhancing the discovery potential for ALPs.

In this paper we provide a first estimate for the experimental sensitivity of {\it Athena} to ALPs. We do so using simulated data for a mock observation of NGC~1275, hosting the radio source 3C~84, which contains the central AGN of the Perseus cluster.
This object was chosen as we have previously used it to place bounds on ALPs using {\it Chandra} data \citep{Berg:2016ese}.

\section{Review of ALP-photon interconversion in clusters}
\label{alps}
An ALP $a$ couples to electromagnetism through the Lagrangian term:
\begin{equation}
\label{ALPphoton}
\qquad L = \frac{1}{4 M}~a~F_{\mu \nu} \tilde{F}^{\mu \nu} =  \frac{1}{M}~a~{\bf E} \cdot {\bf B}~,
\end{equation}
where ${M^{-1} = g_{a\gamma\gamma}}$ parametrises the strength of the interaction,
and ${\bf E}$ and ${\bf B}$ are the electric and magnetic fields.
As their potential and interactions are protected by shift symmetries, ALPs can naturally have very small masses $m_a$. The probability of ALP-photon interaction in the presence of an external magnetic field $\langle B \rangle$ is a standard result \citep{Sikivie:1983ip, Raffelt:1987im}. 

The full analytic expression for the probability of an ALP being converted to a photon after propagating through a single magnetic field domain of length $L$ is:
\begin{equation}
\qquad P_{a \rightarrow \gamma} = \frac{1}{2}\frac{\Theta^2}{1 + \Theta^2}\sin^2 \left( \Delta \sqrt{1 + \Theta^2} \right),
\end{equation}
where
\begin{equation}
\qquad \Theta = 0.28 \Bigg(\frac{B_{\perp}}{1 \mu\rm{G}}\Bigg)\Bigg(\frac{\omega}{1 \, \rm{keV}}\Bigg)\Bigg(\frac{10^{-3} \rm{cm}^{-3}}{n_e}\Bigg)\Bigg(\frac{10^{11} \rm{GeV}}{M}\Bigg),
\end{equation}
\begin{equation}
\qquad \Delta = 0.54 \Bigg(\frac{n_e}{10^{-3} \rm{cm}^{-3}}\Bigg)\Bigg(\frac{L}{10 \, \rm{kpc}}\Bigg)\Bigg(\frac{1 \rm{keV}}{\omega}\Bigg).
\end{equation}
Here $B_{\perp}$ denotes the magnetic field component perpendicular to the ALP wave vector, $\omega$ is the energy and $n_e$ is the electron density. In the limit $\Delta, \Theta \ll 1$, $P \propto B^2 L^2 / M^2$. However when $\Theta < 1$ but $\Delta > 1$, then
$ P \propto \Theta^2 \sin^2 \Delta$.
This probability grows with energy, containing oscillations that are rapid at low energies and broader at higher energies. These
oscillations leave a distinctive imprint on otherwise featureless spectra, and their absence allows us to constrain $g_{a\gamma\gamma}$.

This photon-ALP interconversion is particularly efficient in galaxy clusters (e.g. see ~\citep{0902.2320,1305.3603}).
Clusters have ${\bf B}$ fields of order $\sim \mu$G which extend over megaparsec scales, within which the magnetic field coherence lengths reach tens of kiloparsecs.
The relatively low electron densities ($\sim 10^{-3}\rm{cm}^{-3}$) also implies that it is at X-ray energies that the `sweet spot' of large
 $\Delta$, small $\Theta$, and  quasi-sinusoidal energy-dependent $P_{\gamma \leftrightarrow a}$ is located ~\citep{1304.0989, 1305.3603, 1312.3947, 1509.06748}.

\begin{figure*}
\includegraphics[width=1\textwidth]{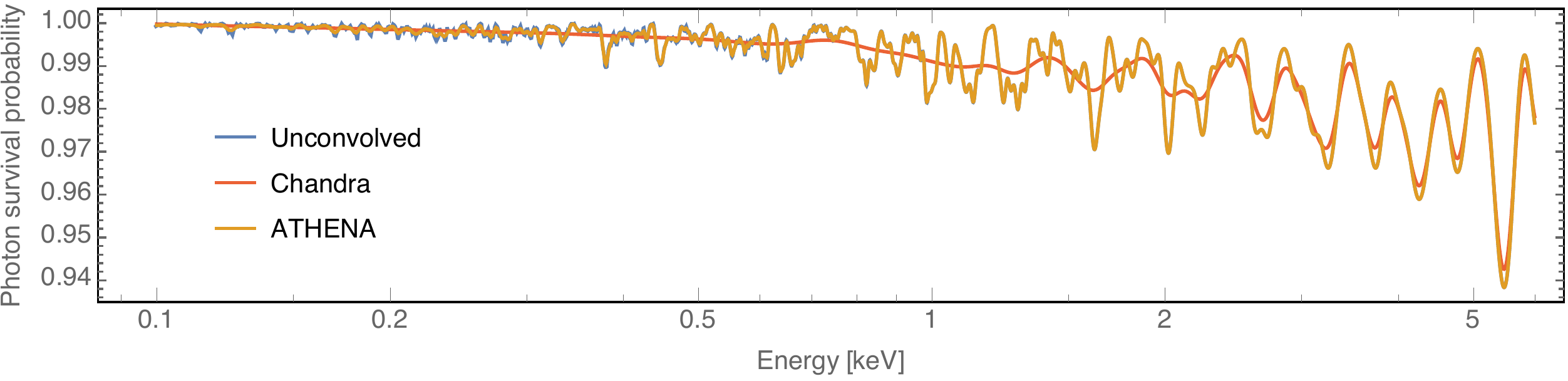} 
\floatbox[{\capbeside\thisfloatsetup{capbesideposition={right,center},capbesidewidth=0.48\textwidth}}]{figure}[\FBwidth]
{\caption{Above---A randomly generated photon survival probability along the line of sight from NGC~1275 to us: unconvolved (blue), convolved with a Gaussian with FWHM 150 eV (a typical energy resolution of {\it Chandra}'s ACIS-I detector (red)) and 2.5 eV for {\it Athena}'s X-IFU detector (orange). A central magnetic field of $B_0 = 25 \mu \rm{G}$ was used, with a radial scaling of $B \sim n_e^{0.7}$, further details in Section~\ref{bounds}. The ALP-photon coupling is ${g_{a\gamma\gamma} = 5 \times 10^{-13} {\rm GeV}^{-1}}$. Small, rapid oscillations at low energies, and larger oscillations at high energies, are generic features of these survival probabilities. At energies <~2~keV {\it Chandra} is unable to resolve oscillations while {\it Athena} performs much better. Left---The same photon survival probabilities, showing the sensitivity of X-IFU to oscillations at low energies.}\label{fig:ALPPhotonConversion}}
{\includegraphics[width=0.48\textwidth]{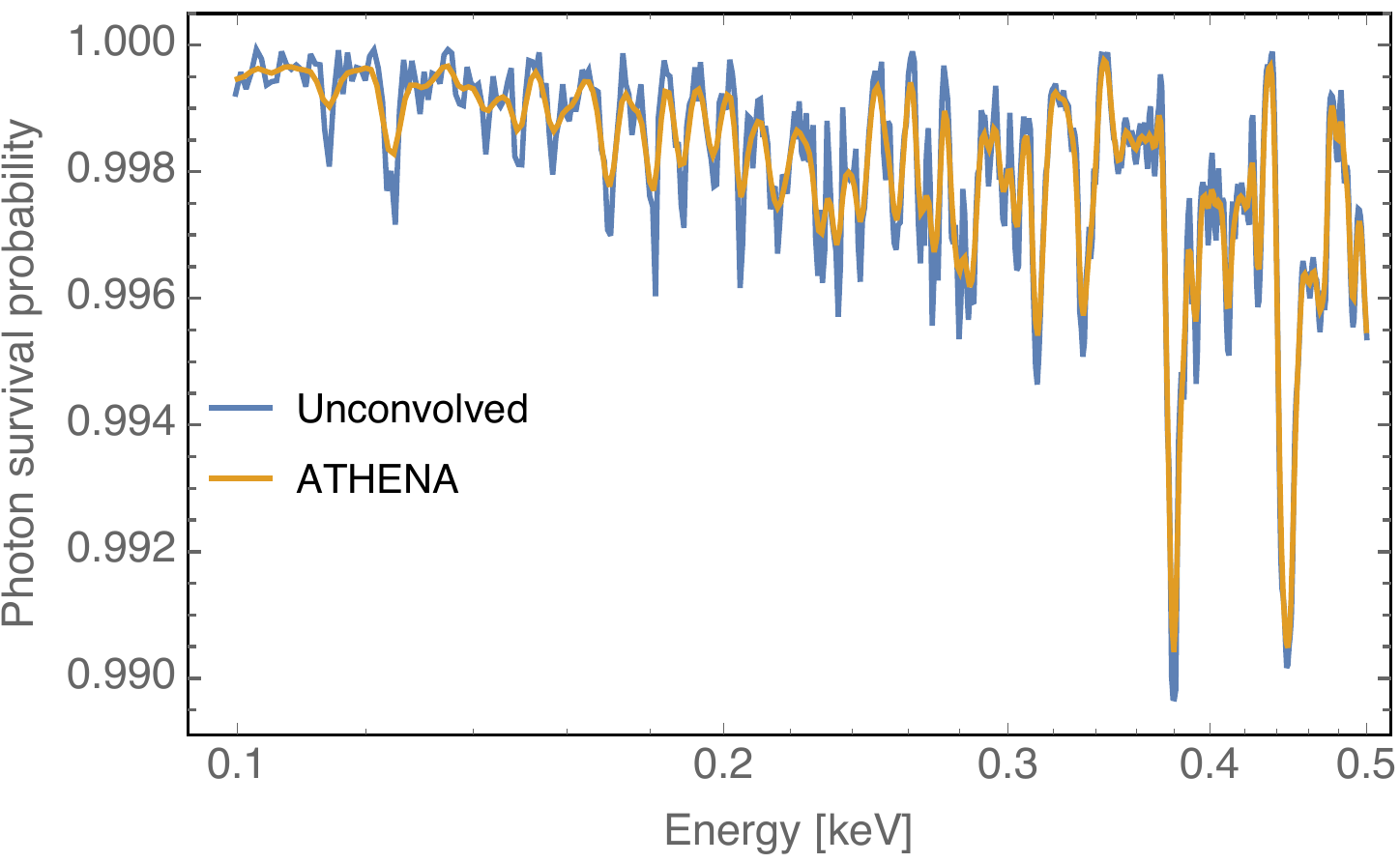}}

\end{figure*}

The 3D structure of intracluster magnetic fields is in general not known and so
the precise form of the survival probability along any single line of sight cannot be determined. Figure~\ref{fig:ALPPhotonConversion} illustrates the energy-dependent survival probability for a photon passing across three hundred domains of a magnetic field, with the direction of the magnetic
field randomised within each domain. The electron density and magnetic field strength in
the model are based on those applicable in the Perseus cluster, but the pattern of smaller, rapid oscillations at low energies and slow oscillations with greater amplitude at high energies is generic.

Active Galactic Nuclei (AGNs) situated in or behind galaxy clusters provide excellent X-ray sources to search for such spectral modulations. One outstanding example is the bright central AGN of the Perseus cluster, at the heart of the galaxy NGC~1275.
Its intrinsic spectrum is well described by an absorbed power law \citep{Churazov:2003hr,Yamazaki, Balmaverde2006, Fabian:2015kua}, and dominates the background cluster emission. The central cluster magnetic field value is estimated at $ \sim 25 \mu$G by \citep{0602622}.

An analysis of archival data of observations of NGC~1275 by the {\it Chandra} and {\it XMM-Newton} satellites was done in \citep{Berg:2016ese}
(see \citep{1603.06978} for a related analysis of NGC~1275 in gamma rays).
Extending methods pioneered in \citep{1304.0989}, the constraint on the ALP-photon coupling ${g_{a\gamma\gamma} \lesssim 1.5 \times 10^{-12}~{\rm GeV}^{-1}}$ was found. For M87, a similar treatment was performed in \citep{Marsh:2017yvc}, finding a
bound ${g_{a\gamma\gamma} \lesssim 1.5 \times 10^{-12}~{\rm GeV}^{-1}}$. An analysis of {\it Chandra} data of other bright point sources in galaxy clusters was conducted in (\citep{Conlon:2017qcw}), deriving bounds of ${g_{a\gamma\gamma} \lesssim 1.5 \times 10^{-12}~{\rm GeV}^{-1}}$ (for the Seyfert galaxy 2E~3140) and ${g_{a\gamma\gamma} \lesssim 2.4 \times 10^{-12}~{\rm GeV}^{-1}}$ (for the AGN NGC~3862).


These bounds all hold for light ALPs with masses $m_a \lesssim 10^{-12} {\rm eV}$. This implies that these methods are not sensitive
to an ordinary QCD axion, which for a photon couplings $g_{a\gamma\gamma} \sim 10^{-12} {\rm GeV}^{-1}$ would typically have $m_a \sim 10^{-3} {\rm eV}$. However, unconventional models for the QCD axion where the photon coupling is significantly enhanced compared to naive expectation may
be constrained using these techniques.

The bounds produced are superior to the bound on light ALPs derived from SN 1987A of $g_{a \gamma \gamma} < 5 \times 10^{-12} \text{GeV}^{-1}$ \citep{Payez:2014xsa}, and are similar to those projected for IAXO in this low mass region \citep{Irastorza:2012qf}. The bounds are also superior to those inferred from the absence of CMB distortions in \mbox{COBE FIRAS} data \citep{Mirizzi:2009nq}, which constrain the product $g_{a \gamma \gamma} B < 10^{-11} \text{GeV}^{-1} \text{nG}$. Here $B$ is the strength of the cosmic magnetic field, which is limited to $B<\text{nG}$.

One major limiting constraint on existing data is the energy resolution of the detectors.
If they exist, ALPs provide oscillatory structure all the way down to the lowest energies.
However, as illustrated in Figure~\ref{fig:ALPPhotonConversion}, detectors with energy resolutions of $\mathcal{O}(100~\rm{eV})$ cannot resolve this structure at lower energies -- but this does become accessible once a resolution of $\mathcal{O}(2.5~\rm{eV})$ is achieved.  We now discuss the future {\it Athena} X-ray observatory, whose greatly enhanced technical capabilities offer improved sensitivity to ALP-photon interconversion.

\section{{\it Athena}}
\label{athena}
The Advanced Telescope for High ENergy Astrophysics (ATHENA) is an ESA mission to explore the Hot and Energetic Universe, due to launch in 2028 \citep{Nandra:2013shg}. The mirror will have a $2~\rm{m}^2$ effective area and a 5~arcsec angular resolution. There are two instruments: the X-ray Integral Field Unit (X-IFU) and the Wide Field Imager (WFI). Here we focus on the former, which will consist of an array of TiAu Transition Edge Sensor (TES) micro-calorimeters sensitive to the energy range 0.2--12~keV \citep{Barret:2016ett}. When operated at a temperature of 50 mK, these can achieve an energy resolution of 2.5~eV below 7~keV \citep{Gottardi:2016cdx}, implying X-IFU will be able to resolve narrow spectral oscillations. A readout time of $\sim10~\mu\rm{s}$ will ensure pileup contamination is minimised. Table~\ref{satellites} contains
a summary of its properties, taken from the {\it Athena} Mission Proposal\footnote{http://www.the-athena-x-ray-observatory.eu/images/AthenaPapers/ \\ The\_Athena\_Mission\_Proposal.pdf}, compared to properties of the {\it Chandra} ACIS-I detector, taken from the {\it Chandra} Proposer's Guide\footnote{http://cxc.harvard.edu/proposer/POG/html/chap6.html}.

\begin{table}

\centering
\begin{tabular}{r|c|c}
& {\it Athena} (X-IFU) & {\it Chandra} (ACIS-I)\\
\hline\hline
Energy range & 0.2--12 keV & 0.3--10 keV\\ \hline
Energy resolution & 2.5 eV & 150 eV\\ 
 at 6 keV & & \\ \hline
Spatial resolution & 5 arcsec & 0.5 arcsec\\ \hline
Time resolution & 10~$\mu$s & 0.2 s\\
& & (2.8~ms single row)\\ \hline
Effective area & 2~m$^2$ @ 1 keV & 600~cm$^2$ @ 1.5 keV \\ \hline
\end{tabular}
\caption{Parameters taken from the {\it Athena} Mission Proposal and the {\it Chandra} Proposer's Guide.}
\label{satellites}
\end{table}

The combination of larger effective area, greatly improved energy resolution and reduced pileup contamination means {\it Athena} has far more potential to detect ALP-induced oscillations than the best current satellites. The aim of this paper is
to make the first quantitative estimate of the extent to which {\it Athena} will be able to improve constraints on
$g_{a\gamma\gamma}$.

\section{Estimate of projected bounds}
\label{bounds}
In terms of estimating bounds on $g_{a\gamma\gamma}$ we use the same method as previously applied with
{\it Chandra} data \citep{Berg:2016ese}. This allows for a direct comparison
between the capabilities of {\it Chandra} and {\it Athena} in terms of placing bounds.

We simulate {\it Athena} observations of NGC~1275, using two models for the photon spectra of the AGN.
The first is a standard spectrum without ALPs, and the second is a model with the same spectrum multiplied with the
photon survival probability distribution as introduced in Section~\ref{alps}. Using simulations of the
X-IFU detector response, we fit spectra generated assuming ALP-photon conversion with the model without ALPs (Model 0) and then we compare this fit to the fit of the spectrum generated without ALPs to the same model.
To allow for the uncertainty in the
magnetic field configuration along the line of sight, we repeat this analysis using many different randomly generated magnetic fields.

The two photon spectra that we model are:
\begin{enumerate}
\item Model 0: An absorbed power law plus thermal background:
\begin{equation}
\qquad F_{0}(E) = (A E^ {- \gamma} + \mathtt{BAPEC}) \times e^{-n_{H} \sigma(E,z)},
\end{equation}
where $A$ and $\gamma$ are the amplitude and index of the power law, $E$ is the energy, $n_H$ is the equivalent hydrogen column, $\sigma(E, z)$ is the photo-electric cross-section at redshift $z$, and $\mathtt{BAPEC}$ is the standard plasma thermal emission model.
\item Model 1: An absorbed power law plus thermal background, multiplied by a table of survival probabilities for photons of different energies:
\begin{equation}
F_{1}(E, {\bf B}) = (A E^ {- \gamma} + \mathtt{BAPEC}) \times e^{-n_{H} \sigma(E,z)} \times P_{\gamma \to \gamma} (E (1 + z), { \bf B}, g_{a \gamma \gamma})~.
\end{equation}
\end{enumerate}
The index of the power law was set based on the best fit value from the cleanest {\it Chandra} observations of NGC~1275, and its normalisation was determined based on the {\it Hitomi} 230~ks observation of Perseus in 2016 \citep{Aharonian:2016gzq}. As the AGN in 2016 was
roughly twice as bright as in 2009 and it has previously exhibited large historical variation \citep{Fabian:2015kua}, it may be again much brighter (or dimmer) in 2028, which would affect both the contrast
against the cluster background and also the observation time required to achieve a certain constraint on $g_{a \gamma \gamma}$.

The 2016 {\it Hitomi} observation also constrained the temperature, abundances and velocity dispersion of the cluster thermal emission to a high degree of accuracy \citep{Aharonian:2016gzq}.
For the spectral shape of the cluster background, we used the single-temperature $\mathtt{bapec}$ model that was a good fit
to the {\it Hitomi} spectrum across its field of view. While this single-temperature model is unlikely to be a good fit for the background
contiguous to the AGN, it represents a useful proxy for the actual background that can only be determined at the time.
The normalisation of the background was set by extracting a circular region
of the cluster emission close to the AGN from the {\it Chandra} observations, of radius equal to the angular resolution of {\it Athena}, and determining the best fit. All model parameters are shown in Table \ref{parameters}.
\begin{table}
\centering
\begin{tabular}{l|l|c|c}
Model & parameter & symbol & value\\ \hline \hline
$\mathtt{zwabs}$ & nH column density & $n_H$ & $0.24 \times 10^{22} \rm{cm}^{-2}$\\
 & redshift & z & 0.0176\\ \hline
$\mathtt{powerlaw}$ & index & $\gamma$ & 1.8\\
 & normalisation & A & $9 \times 10^{-3}$\\ \hline
$\mathtt{bapec}$ & temperature & kT & 3.48 keV\\
 & abundances & & 0.54 solar\\
 & velocity dispersion & v & 178 m \,${\rm s}^{-1}$\\
& normalisation & N & $9 \times 10^{-4}$
\end{tabular}
\caption{Parameters of the absorbed power law describing the spectrum of NGC~1275, and the thermal model of the cluster background.}
\label{parameters}
\end{table}

As for the study with {\it Chandra}, we take the central magnetic field value as $B_0\sim 25\mu $G, following \citep{0602622}. We also
assume that $B$ decreases with radius as $B \propto n_{e}^{0.7}$.
As there is not a direct measurement of the power spectrum and coherence length for the Perseus magnetic field, we base the model on those inferred for the cool core cluster A2199 \citep{Vacca:2012up}.

The electron density $n_{e}$ has the radial distribution found in \citep{Churazov:2003hr}:
\begin{equation}
\qquad n_{e} (r) = \frac{3.9 \times 10^{-2}}{ [ 1 + (\frac{r}{80 \, {\rm kpc}})^2]^{1.8}} +  \frac{4.05 \times 10^{-3}}{ [ 1 + (\frac{r}{280 \, {\rm kpc}})^2]^{0.87}} \, {\rm cm}^{-3}.
\end{equation}

The magnetic field is generated over 300 domains, whose lengths are drawn from a Pareto distribution between $3.5~\rm{kpc}$ and $10~\rm{kpc}$ with power 2.8. In each domain the magnetic field and electron density are constant, with a random direction of ${\bf B}$. We then
calculate the survival probability of a photon passing through this region, as described in \citep{1312.3947}.

The simulations were performed using the Simulation of X-ray Telescopes ($\mathtt{SIXTE}$) code, a multi-instrument simulation package. It aims to offer an end-to-end simulation, i.e. the full detector chain from the source to the final data. It models the telescope's vignetting, ARF and PSF, and X-IFU's response, event reconstruction and pileup \citep{2014SPIE.9144E..5XW}.

\begin{figure}
\includegraphics[width=1.0\textwidth]{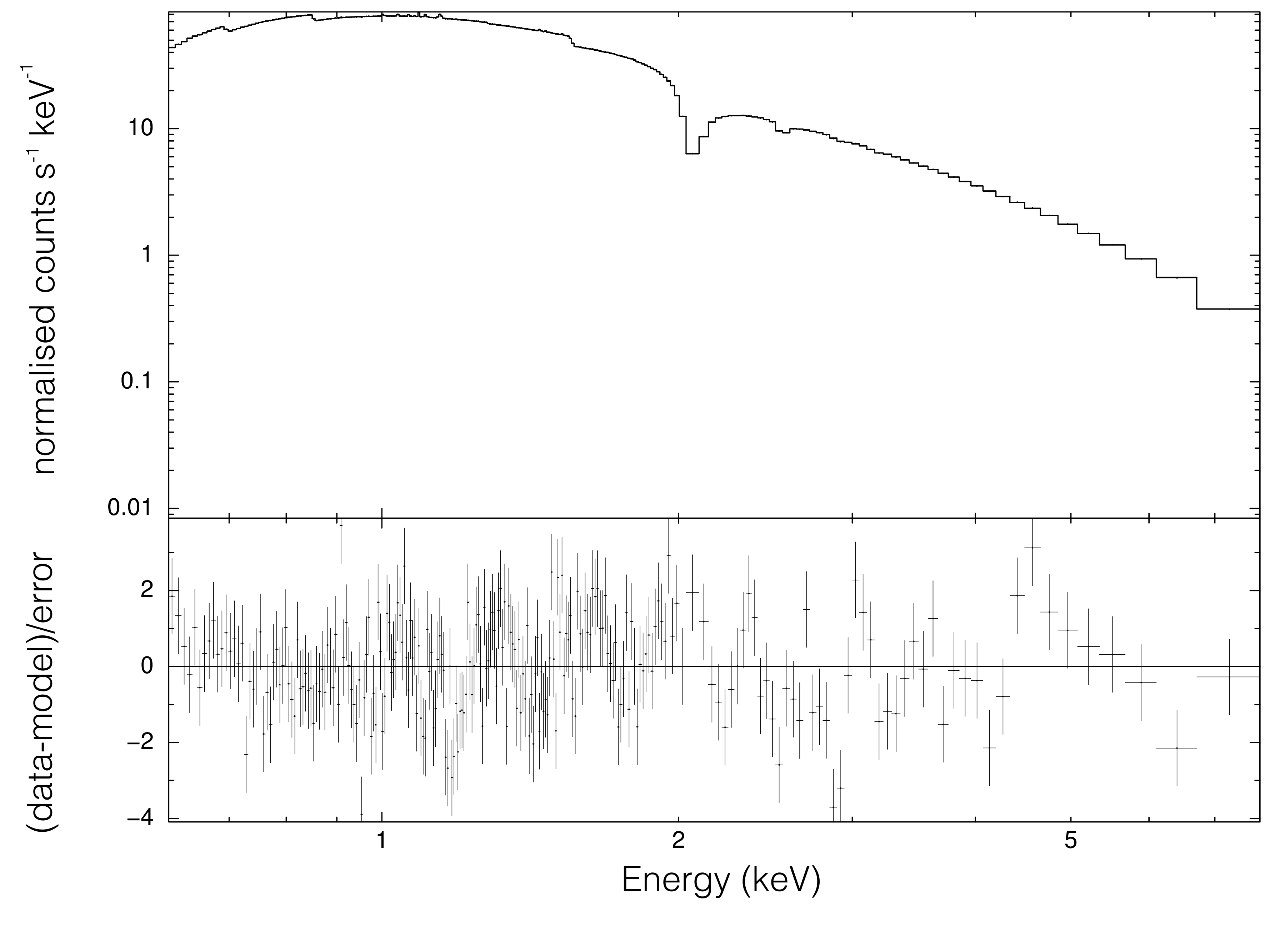}
\caption{A simulated 200~ks dataset for NGC~1275 with ${g_{a \gamma \gamma} = 3 \times 10^{-13} \rm{GeV}^{-1}}$, and its fit to Model 0. The characteristic ALP-induced modulations are apparent.}
\label{spectrum}
\end{figure}

The spectrum of NGC~1275, and the cluster background, were modelled in $\mathtt{XSPEC}$\footnote{https://heasarc.gsfc.nasa.gov/xanadu/xspec/manual/manual.html} as an absorbed power law plus a thermal component, $\mathtt{zwabs*(powerlaw + bapec)}$. This spectrum, either multiplied with the photon survival probabilities or not, was converted to the SIMPUT\footnote{http://hea-www.harvard.edu/heasarc/formats/simput-1.0.0.pdf} file format using the command $\mathtt{simputfile}$. The mirror and detector response were modelled with $\mathtt{xifupipeline}$, using the ARF file $\mathtt{athena\_xifu\_1469\_onaxis\_pitch249um\_v20160401.arf}$ and the RMF file $\mathtt{athena\_xifu\_rmf\_v20160401.rmf}$. This generated an event FITS file, which was then converted into a PHA file using $\mathtt{makespec}$. We produced a fit to this spectrum in $\mathtt{XSPEC}$, using the Levenberg-Marquardt fitting method to calculate the reduced $\chi^2$. Figure~\ref{spectrum} shows one simulation for ${g_{a \gamma \gamma} = 3 \times 10^{-13} {\rm GeV}^{-1}}$ and its fit to an absorbed power law.

We use the following procedure to determine whether a particular value of $g_{a \gamma \gamma}$ is excluded: we varied the ALP-photon coupling $g_{a \gamma \gamma}$ from $g_{a \gamma \gamma} = 5 \times 10^{-13} \, \text{GeV}^{-1}$ to $g_{a \gamma \gamma} = 1\times 10^{-13} \, \text{GeV}^{-1}$, with stepsize $0.5 \times 10^{-13} \, \text{GeV}^{-1}$. As the bound is dependent on uncertainties in the magnetic field strength of a factor of 2, and we are only using simulated data, we do not consider step sizes smaller than this. For each $g_{a \gamma \gamma}$:
\begin{enumerate}
\item Generate 50 configurations of the magnetic field $B_i$.
\item Use the $B_i$ to calculate the survival probability $P_{\gamma \rightarrow \gamma}$ along the line of sight for different photon energies (as done in \citep{1312.3947}). We calculate for 8000 equally spaced photon energies in the range 0.01--10~keV.
\item Combine each $P_{\gamma \rightarrow \gamma}$ with the AGN spectrum.
\item Generate 10 fake PHAs for each spectrum, providing 500 fake data samples in total.
\item Fit the fake data to Model 0, and calculate the reduced chi-squareds $\chi_1^2$.
\item Generate 100 fake PHAs based on Model 0, and compute the average of their reduced chi-squareds $\chi_0^2$.
Assuming the absence of ALPs, this represents the expected quality of the fit to the single real data set. If the actual data is a poor fit
for some reason, then this will weaken the level of the resulting bounds that we can produce.
\item Determine the percentage of fake data sets that have a reduced chi-squared $\chi_1^2 < \rm{max}(\langle \chi_0^2 \rangle,1)$. If this is true for fewer than 5 per cent of the data sets, the value of $g_{a \gamma \gamma}$ is excluded at 95 per cent confidence.
\end{enumerate}

\begin{figure}
\includegraphics[width=1.0\textwidth]{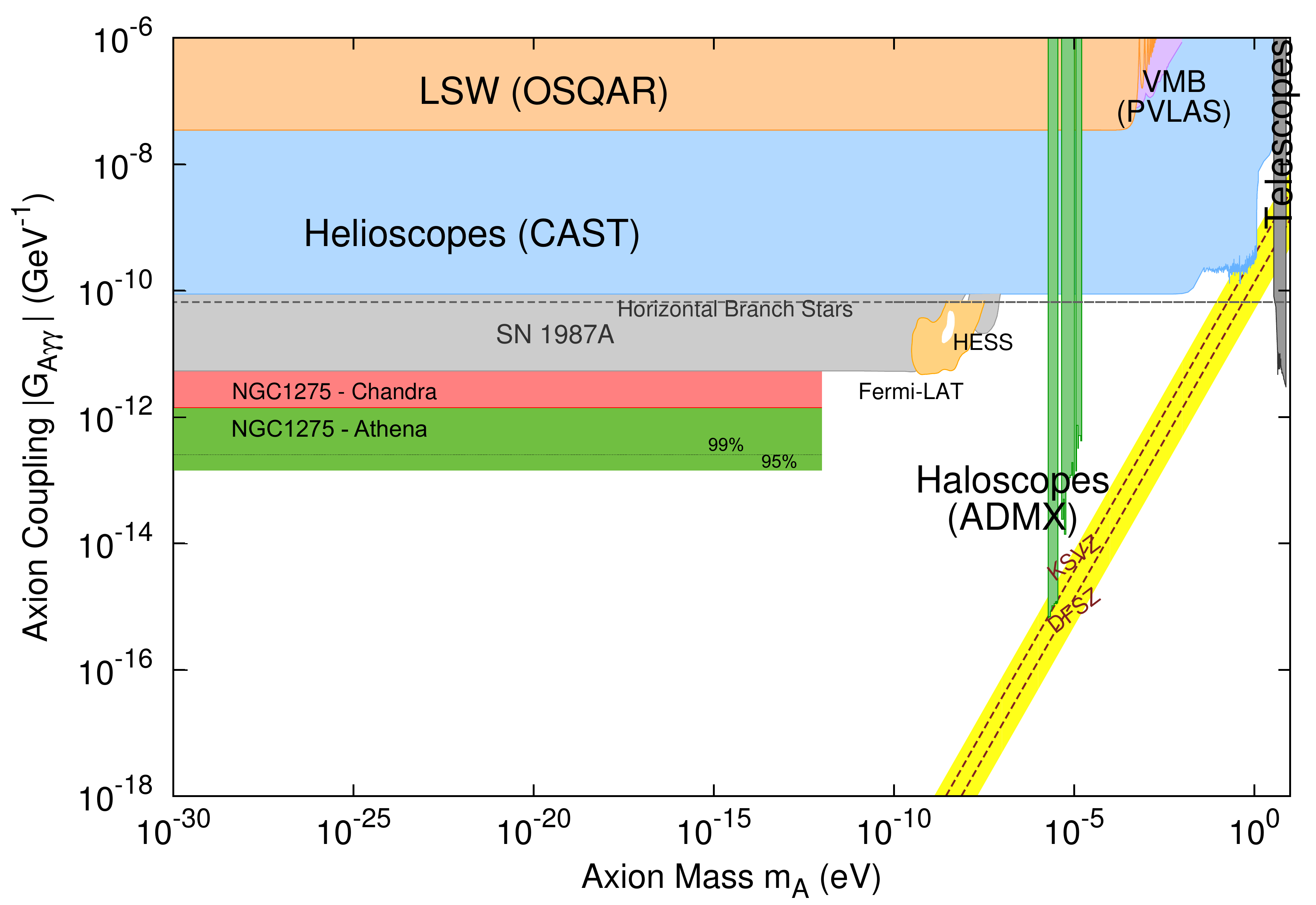}
\caption{Overview of exclusion limits on axion couplings vs mass. For axion masses $m_a \sim 10^{-12} {\rm eV}$ then ALP-photon conversion
can enter a resonant regime, with the potential of stronger bounds around this critical mass. We do not perform a detailed study of the resonant
regime in this work and focus only on the low-mass region. Full references can be found in the Particle Data Group review on {\it Axions and other similar particles} \citep{Patrignani:2016xqp}.}\label{fig:ExclusionLimit}
\end{figure}

For a simulation of 200~ks of data with the nominal mirror configuration, we derive a projected
bound of ${g_{a\gamma\gamma} \lesssim 1.5 \times 10^{-13}~ \rm{GeV}^{-1}}$ at 95 per cent confidence and of ${g_{a\gamma\gamma} \lesssim 2.5 \times 10^{-13}~ \rm{GeV}^{-1}}$ at 99\% confidence, as shown in Fig.~\ref{fig:ExclusionLimit} alongside published data limits. This represents an order of magnitude improvement over the bound derived from the 200~ks of Chandra ACIS-I observations in \citep{Berg:2016ese}. We also find that even
a short 10~ks observation will lead to an improved bound of ${g_{a\gamma\gamma} \lesssim 4.5 \times 10^{-13}~ \rm{GeV}^{-1}}$.

These bounds are substantially better than any current experimental or astrophysical bound, and also go beyond the capabilities of IAXO for ultralight ALP masses. The proposed DM haloscope ABRACADABRA has the potential to explore $g_{a \gamma \gamma}$ down to $10^{-17} \text{GeV}^{-1}$ for $m_a \in [10^{-14},10^{-6}] \text{eV}$  \citep{Kahn:2016aff}, if ALPs constitute the Dark Matter. The existence of ALP-induced oscillations in galaxy clusters is independent of this. Proposed CMB experiments such as PIXIE \citep{Kogut:2011xw} and PRISM \citep{Andre:2013afa} could produce a constraint $g_{a \gamma \gamma} B < 10^{-16} \text{GeV}^{-1} \text{nG}$ which might be competitive with bounds from galaxy clusters if the cosmic magnetic field is close enough to saturation $\sim \text{nG}$ \citep{Tashiro:2013yea}. Black hole superradiance also offers tentative constraints ALPs on in the mass range $m_a \in [10^{-14},10^{-10}] \text{eV}$, depending on measurements of black hole spin \citep{Arvanitaki:2016qwi}.

\section{Conclusion}
\label{conclusion}
AGNs situated in galaxy clusters are excellent targets to search for ALP-photon interconversion. {\it Athena}'s groundbreaking new technology will be able to resolve AGN spectra very precisely. The bound ${g_{a\gamma\gamma} \lesssim 1.5 \times 10^{-13}~ \rm{GeV}^{-1}}$ derived from simulations of 200~ks observations is an order of magnitude improvement over the bounds from current generation satellites. For the mass range ${m_a \lesssim 10^{-12}~\rm{eV}}$, it will also be far better than the bounds obtainable from future experimental searches such as IAXO.

We stress that this is only a first estimate of  the sensitivity of {\it Athena} to ALP-induced modulations. The final sensitivity will depend on the capabilities of the finished satellite, the brightness of the AGN in 2028 and the quality of the actual data.
Telescopes such as the Square Kilometre Array (SKA) are likely to reduce the uncertainties in the magnetic field model \citep{Braun:2015zta}, allowing for greater precision in $g_{a \gamma \gamma}$ bounds calculations by the time {\it Athena} launches. However, we have demonstrated that {\it Athena} will certainly improve bounds on $g_{a \gamma \gamma}$ substantively, and that X-ray astronomy will continue to be at the forefront of ultralight ALP searches in the coming decades.

\section*{Acknowledgements}

This project is funded in part by the European Research Council starting grant `Supersymmetry Breaking in String Theory' (307605). Both Francesca Day and Nicholas Jennings are also funded by STFC.\\
\bsp
\bibliographystyle{mnras}
\bibliography{Athenabib}

\label{lastpage}
\end{document}